\documentclass[reprint, superscriptaddress, prl]{revtex4-1}

\usepackage{amsmath,amssymb}
\usepackage{graphicx}
\usepackage{bm}
\usepackage{times}
\usepackage{lineno}
\usepackage{textcomp, gensymb}
\usepackage[utf8]{inputenc}
\usepackage{hyperref}

\newcommand{\nuebar}{\ensuremath{\overline{\nu}_{e}}}
\newcommand{\uFive}{$^{235}$U}
\newcommand{\uEight}{$^{238}$U}
\newcommand{\pNine}{$^{239}$Pu}
\newcommand{\pOne}{$^{241}$Pu}
\newcommand{\erec}{E$_{rec}$}

\begin{document}

\title{Measurement of the Antineutrino Spectrum from \uFive{} Fission at HFIR with PROSPECT} 

\affiliation{Brookhaven National Laboratory, Upton, NY, USA}
\affiliation{Department of Physics, Drexel University, Philadelphia, PA, USA}
\affiliation{George W.\,Woodruff School of Mechanical Engineering, Georgia Institute of Technology, Atlanta, GA, USA}
\affiliation{Department of Physics, Illinois Institute of Technology, Chicago, IL, USA}
\affiliation{Nuclear and Chemical Sciences Division, Lawrence Livermore National Laboratory, Livermore, CA, USA}
\affiliation{Department of Physics, Le Moyne College, Syracuse, NY, USA}
\affiliation{National Institute of Standards and Technology, Gaithersburg, MD, USA}
\affiliation{High Flux Isotope Reactor, Oak Ridge National Laboratory, Oak Ridge, TN, USA}
\affiliation{Physics Division, Oak Ridge National Laboratory, Oak Ridge, TN, USA}
\affiliation{Department of Physics, Temple University, Philadelphia, PA, USA}
\affiliation{Department of Physics and Astronomy, University of Tennessee, Knoxville, TN, USA}
\affiliation{Institute for Quantum Computing and Department of Physics and Astronomy, University of Waterloo, Waterloo, ON, Canada}
\affiliation{Department of Physics, University of Wisconsin, Madison, Madison, WI, USA}
\affiliation{Physical Sciences Laboratory, University of Wisconsin, Madison, Madison, WI, USA}
\affiliation{Department of Physics, College of William and Mary, Williamsburg, VA, USA}
\affiliation{Wright Laboratory, Department of Physics, Yale University, New Haven, CT, USA}

\author{J.\,Ashenfelter}
\affiliation{Wright Laboratory, Department of Physics, Yale University, New Haven, CT, USA}
\author{A.\,B.\,Balantekin}
\affiliation{Department of Physics, University of Wisconsin, Madison, Madison, WI, USA}
\author{H.\,R.\,Band}
\affiliation{Wright Laboratory, Department of Physics, Yale University, New Haven, CT, USA}
\author{C.\,D.\,Bass}
\affiliation{Department of Physics, Le Moyne College, Syracuse, NY, USA}
\author{D.\,E.\,Bergeron}
\affiliation{National Institute of Standards and Technology, Gaithersburg, MD, USA}
\author{D.\,Berish}
\affiliation{Department of Physics, Temple University, Philadelphia, PA, USA}
\author{N.\,S.\,Bowden}
\affiliation{Nuclear and Chemical Sciences Division, Lawrence Livermore National Laboratory, Livermore, CA, USA}
\author{J.\,P.\,Brodsky}
\affiliation{Nuclear and Chemical Sciences Division, Lawrence Livermore National Laboratory, Livermore, CA, USA}
\author{C.\,D.\,Bryan}
\affiliation{High Flux Isotope Reactor, Oak Ridge National Laboratory, Oak Ridge, TN, USA}
\author{J.\,J.\,Cherwinka}
\affiliation{Physical Sciences Laboratory, University of Wisconsin, Madison, Madison, WI, USA}
\author{T.\,Classen}
\affiliation{Nuclear and Chemical Sciences Division, Lawrence Livermore National Laboratory, Livermore, CA, USA}
\author{A.\,J.\,Conant}
\affiliation{George W.\,Woodruff School of Mechanical Engineering, Georgia Institute of Technology, Atlanta, GA, USA}
\author{A.\,A.\,Cox}
\affiliation{Institute for Quantum Computing and Department of Physics and Astronomy, University of Waterloo, Waterloo, ON, Canada}
\author{D.\,Davee}
\affiliation{Department of Physics, College of William and Mary, Williamsburg, VA, USA}
\author{D.\,Dean}
\affiliation{Physics Division, Oak Ridge National Laboratory, Oak Ridge, TN, USA}
\author{G.\,Deichert}
\affiliation{High Flux Isotope Reactor, Oak Ridge National Laboratory, Oak Ridge, TN, USA}
\author{M.\,V.\,Diwan}
\affiliation{Brookhaven National Laboratory, Upton, NY, USA}
\author{M.\,J.\,Dolinski}
\affiliation{Department of Physics, Drexel University, Philadelphia, PA, USA}
\author{A.\,Erickson}
\affiliation{George W.\,Woodruff School of Mechanical Engineering, Georgia Institute of Technology, Atlanta, GA, USA}
\author{M.\,Febbraro}
\affiliation{Physics Division, Oak Ridge National Laboratory, Oak Ridge, TN, USA}
\author{B.\,T.\,Foust}
\affiliation{Wright Laboratory, Department of Physics, Yale University, New Haven, CT, USA}
\author{J.\,K.\,Gaison}
\affiliation{Wright Laboratory, Department of Physics, Yale University, New Haven, CT, USA}
\author{A.\,Galindo-Uribarri}
\affiliation{Physics Division, Oak Ridge National Laboratory, Oak Ridge, TN, USA}
\affiliation{Department of Physics and Astronomy, University of Tennessee, Knoxville, TN, USA}
\author{C.\,E.\,Gilbert}
\affiliation{Physics Division, Oak Ridge National Laboratory, Oak Ridge, TN, USA}
\affiliation{Department of Physics and Astronomy, University of Tennessee, Knoxville, TN, USA}
\author{K.\,E.\,Gilje}
\affiliation{Department of Physics, Illinois Institute of Technology, Chicago, IL, USA}
\author{B.\,T.\,Hackett}
\affiliation{Physics Division, Oak Ridge National Laboratory, Oak Ridge, TN, USA}
\affiliation{Department of Physics and Astronomy, University of Tennessee, Knoxville, TN, USA}
\author{S.\,Hans}
\thanks{Also at: Department of Chemistry and Chemical Technology, Bronx Community College, Bronx, NY, USA.}
\affiliation{Brookhaven National Laboratory, Upton, NY, USA}
\author{A.\,B.\,Hansell}
\affiliation{Department of Physics, Temple University, Philadelphia, PA, USA}
\author{K.\,M.\,Heeger}
\affiliation{Wright Laboratory, Department of Physics, Yale University, New Haven, CT, USA}
\author{J.\,Insler}
\affiliation{Department of Physics, Drexel University, Philadelphia, PA, USA}
\author{D.\,E.\,Jaffe}
\affiliation{Brookhaven National Laboratory, Upton, NY, USA}
\author{X.\,Ji}
\affiliation{Brookhaven National Laboratory, Upton, NY, USA}
\author{D.\,C.\,Jones}
\affiliation{Department of Physics, Temple University, Philadelphia, PA, USA}
\author{O.\,Kyzylova}
\affiliation{Department of Physics, Drexel University, Philadelphia, PA, USA}
\author{C.\,E.\,Lane}
\affiliation{Department of Physics, Drexel University, Philadelphia, PA, USA}
\author{T.\,J.\,Langford}
\thanks{Also at: Yale Center for Research Computing, Yale University, New Haven CT 06520}
\affiliation{Wright Laboratory, Department of Physics, Yale University, New Haven, CT, USA}
\author{J.\,LaRosa}
\affiliation{National Institute of Standards and Technology, Gaithersburg, MD, USA}
\author{B.\,R.\,Littlejohn}
\affiliation{Department of Physics, Illinois Institute of Technology, Chicago, IL, USA}
\author{X.\,Lu}
\affiliation{Physics Division, Oak Ridge National Laboratory, Oak Ridge, TN, USA}
\affiliation{Department of Physics and Astronomy, University of Tennessee, Knoxville, TN, USA}
\author{D.\,A.\,Martinez\,Caicedo}
\affiliation{Department of Physics, Illinois Institute of Technology, Chicago, IL, USA}
\author{J.\,T.\,Matta}
\affiliation{Physics Division, Oak Ridge National Laboratory, Oak Ridge, TN, USA}
\author{R.\,D.\,McKeown}
\affiliation{Department of Physics, College of William and Mary, Williamsburg, VA, USA}
\author{M.\,P.\,Mendenhall}
\affiliation{Nuclear and Chemical Sciences Division, Lawrence Livermore National Laboratory, Livermore, CA, USA}
\author{J.\,M.\,Minock}
\affiliation{Department of Physics, Drexel University, Philadelphia, PA, USA}
\author{P.\,E.\,Mueller}
\affiliation{Physics Division, Oak Ridge National Laboratory, Oak Ridge, TN, USA}
\author{H.\,P.\,Mumm}
\affiliation{National Institute of Standards and Technology, Gaithersburg, MD, USA}
\author{J.\,Napolitano}
\affiliation{Department of Physics, Temple University, Philadelphia, PA, USA}
\author{R.\,Neilson}
\affiliation{Department of Physics, Drexel University, Philadelphia, PA, USA}
\author{J.\,A.\,Nikkel}
\affiliation{Wright Laboratory, Department of Physics, Yale University, New Haven, CT, USA}
\author{D.\,Norcini}
\affiliation{Wright Laboratory, Department of Physics, Yale University, New Haven, CT, USA}
\author{S.\,Nour}
\affiliation{National Institute of Standards and Technology, Gaithersburg, MD, USA}
\author{D.\,A.\,Pushin}
\affiliation{Institute for Quantum Computing and Department of Physics and Astronomy, University of Waterloo, Waterloo, ON, Canada}
\author{X.\,Qian}
\affiliation{Brookhaven National Laboratory, Upton, NY, USA}
\author{E.\,Romero-Romero}
\affiliation{Physics Division, Oak Ridge National Laboratory, Oak Ridge, TN, USA}
\affiliation{Department of Physics and Astronomy, University of Tennessee, Knoxville, TN, USA}
\author{R.\,Rosero}
\affiliation{Brookhaven National Laboratory, Upton, NY, USA}
\author{D.\,Sarenac}
\affiliation{Institute for Quantum Computing and Department of Physics and Astronomy, University of Waterloo, Waterloo, ON, Canada}
\author{P.\,T.\,Surukuchi}
\affiliation{Department of Physics, Illinois Institute of Technology, Chicago, IL, USA}
\author{A.\,B.\,Telles}
\affiliation{Wright Laboratory, Department of Physics, Yale University, New Haven, CT, USA}
\author{M.\,A.\,Tyra}
\affiliation{National Institute of Standards and Technology, Gaithersburg, MD, USA}
\author{R.\,L.\,Varner}
\affiliation{Physics Division, Oak Ridge National Laboratory, Oak Ridge, TN, USA}
\author{B.\,Viren}
\affiliation{Brookhaven National Laboratory, Upton, NY, USA}
\author{C.\,White}
\affiliation{Department of Physics, Illinois Institute of Technology, Chicago, IL, USA}
\author{J.\,Wilhelmi}
\affiliation{Department of Physics, Temple University, Philadelphia, PA, USA}
\author{T.\,Wise}
\affiliation{Wright Laboratory, Department of Physics, Yale University, New Haven, CT, USA}
\author{M.\,Yeh}
\affiliation{Brookhaven National Laboratory, Upton, NY, USA}
\author{Y.-R.\,Yen}
\affiliation{Department of Physics, Drexel University, Philadelphia, PA, USA}
\author{A.\,Zhang}
\affiliation{Brookhaven National Laboratory, Upton, NY, USA}
\author{C.\,Zhang}
\affiliation{Brookhaven National Laboratory, Upton, NY, USA}
\author{X.\,Zhang}
\affiliation{Department of Physics, Illinois Institute of Technology, Chicago, IL, USA}

\collaboration{The PROSPECT Collaboration}
\email{prospect.collaboration@gmail.com}

\date{\today}

\begin{abstract}
This Letter reports the first measurement of the \uFive{} \nuebar{} energy spectrum by PROSPECT, the Precision Reactor Oscillation and Spectrum experiment, operating 7.9 m from the 85\,MW$_{\mathrm{th}}$ highly-enriched uranium (HEU) High Flux Isotope Reactor. 
With a surface-based, segmented detector, PROSPECT has observed 31678\,$\pm$\,304 (stat.) \nuebar{}-induced inverse beta decays (IBD), the largest sample from HEU fission to date, 99\,\% of which are attributed to \uFive{}.  
Despite broad agreement, comparison of the Huber \uFive{} model to the measured spectrum produces a $\chi^2/ndf = 51.4/31$, driven primarily by deviations in two localized energy regions. 
The measured \uFive{} spectrum shape is consistent with a deviation relative to prediction equal in size to that observed at low-enriched uranium power reactors in the \nuebar{} energy region of 5-7\,MeV.

\end{abstract}

\maketitle

  
Reactor \nuebar{} experiments have been central to the understanding of neutrinos, including the first observation of \nuebar{}~\cite{Cowan:1992xc}, the discovery of \nuebar{} oscillations~\cite{Eguchi:2002dm}, observation of \nuebar{} produced within the Earth~\cite{Araki:2005qa}, and the measurement of the neutrino mixing angle $\theta_{13}$~\cite{An:2012eh,Ahn:2012nd,Abe:2011fz}.
Most of these experiments were located at low-enriched uranium (LEU) nuclear power reactors where more than 99\,\% of emitted \nuebar{} come from the beta decay of fission products of four isotopes  (\uFive{}, \uEight{}, \pNine{}, and \pOne{}). 
At power reactors, the emitted \nuebar{} flux and spectrum evolve over time as the isotopic composition changes in the fuel cycle.
Comparisons between theoretical predictions and experimental results reveal a $\sim$6\,\% global flux deficit~\cite{Mueller:2011nm, Mention:2011rk,Huber:2011wv,Fallot:2012jv} and disagreement of the energy spectrum~\cite{An:2015nua, Abe:2014bwa, Seo:2014xei, Ko:2016owz} and flux-evolution~\cite{An:2017osx, RENO:2018pwo}.  
Explanations for these possibly independent phenomena may lie in the complex nuclear physics of reactors~\cite{Hayes:2013wra, Dwyer:2014eka, Hayes:2015yka, Wang:2016rqh, Wang:2017htp, Sonzogni:2017wxy, Gebre:2017vmm, Littlejohn:2018hqm}, physics beyond the Standard Model such as eV-scale sterile neutrinos~\cite{Mention:2011rk}, or both~\cite{Giunti:2017yid, Dentler:2017tkw, Giunti:2019qlt}.
New experiments at compact-core, highly enriched uranium (HEU) research reactors enable short baseline searches for sterile neutrino oscillations and the measurement of the nearly time-independent emission of \nuebar{} from \uFive{} fission~\cite{Allemandou:2018vwb, Abreu:2017bpe, Ashenfelter:2018zdm}.
PROSPECT has recently reported a search for sterile neutrinos at the High Flux Isotope Reactor (HFIR)~\cite{Ashenfelter:2018iov}. 
This Letter reports the first measurement of the \nuebar{} energy spectrum from HFIR by the PROSPECT experiment and the highest-statistics \uFive{} spectral measurement since the ILL experiment observed $\sim$5000 \nuebar{} candidates in 1981~\cite{Kwon:1981ua}. 
 

Located at Oak Ridge National Laboratory, HFIR is an 85~megawatt thermal (MW$\mathrm{_{th}}$) HEU research reactor.
The cylindrical reactor core (diameter: 0.435\,m, height: 0.508\,m) contains 93\,\% \uFive{} enriched fuel, leading to a $\sim$99\,\% \uFive{} fission fraction. 
Each 24-day operating cycle uses fresh fuel, minimizing \pNine{} and \pOne{} production. 
The PROSPECT detector is deployed in a ground-level room at a center-to-center distance of (7.9\,$\pm$\,0.1)\,m from the reactor core.
The core center is located 40\degree{} below the horizontal and the surrounding building provides less than one meter-water-equivalent of concrete overburden.

PROSPECT uses inverse beta decay (IBD), $\nuebar{} + p \rightarrow \beta^{+} + n$, to detect \nuebar{} within a 4-tonne $^{6}$Li-loaded liquid scintillator ($^{6}$LiLS) target divided into an 11x14 array of optically isolated rectangular segments~\cite{Ashenfelter:2018cli, Ashenfelter:2019iqj, Ashenfelter:2019lbf}.
The measured energy of $\beta^{+}$ ionization and annihilation, or prompt signal, carries the \nuebar{} energy information.
The delayed neutrons principally capture on $^{6}$Li (nLi) with an average time separation of $\sim$50\,$\micro$s.
This prompt-delay coincident pair identifies IBD-like events.
Each 14.5~cm $\times$ 14.5~cm $\times$ 117.6~cm segment is read out on both ends by 12.7~cm photomultiplier tubes (PMTs). 
Segments are rotated by 5$\degree$ to create space for hollow plastic support rods at each corner, allowing radioactive sources and optical calibration inserts to be deployed adjacent to all fiducial segments. 

Waveform digitizers (WFDs, 250\,MHz, 14-bit) record signals generated by scintillation light collected by PMTs.
The triggering scheme balances overall throughput with the need to capture small energy depositions from Compton scattering of 511\,keV annihilation $\gamma$-rays.
Observation of coincident signals in the two PMTs of any segment (combining to $\sim$150\,keV deposition threshold) triggers waveform acquisition of all WFD channels.
PMT signals that exceed a zero-suppression threshold are stored for offline analysis.  
Individual pulses are integrated to determine the amount of light observed by each PMT.
The time-separation and charge-ratio of the two PMT signals from each segment are combined to determine the interaction position (z) along the segment length.
The energy deposition per segment is determined from a position-corrected PE count.
The relative energy scale of each segment is determined from fitting the nLi capture peak.
Depositions occurring within $\sim$20~ns are summed to produce the total reconstructed visible energy (\erec{}).
Variations in hardware digitization thresholds and light collection along segments are controlled by the application of a segment-wise 85~keV analysis threshold resulting in uniform event acceptance across the entire detector volume.  
A metric for particle identification, Pulse Shape Discrimination (PSD), is determined from the ratio of the PMT pulse tail to the total pulse integral, effectively separating interactions from heavy and light charged particles. 
Further information on the detector design and data acquisition are given in Ref.~\cite{Ashenfelter:2018zdm}.

\begin{figure}
    \centering
    \includegraphics[width=0.45\textwidth]{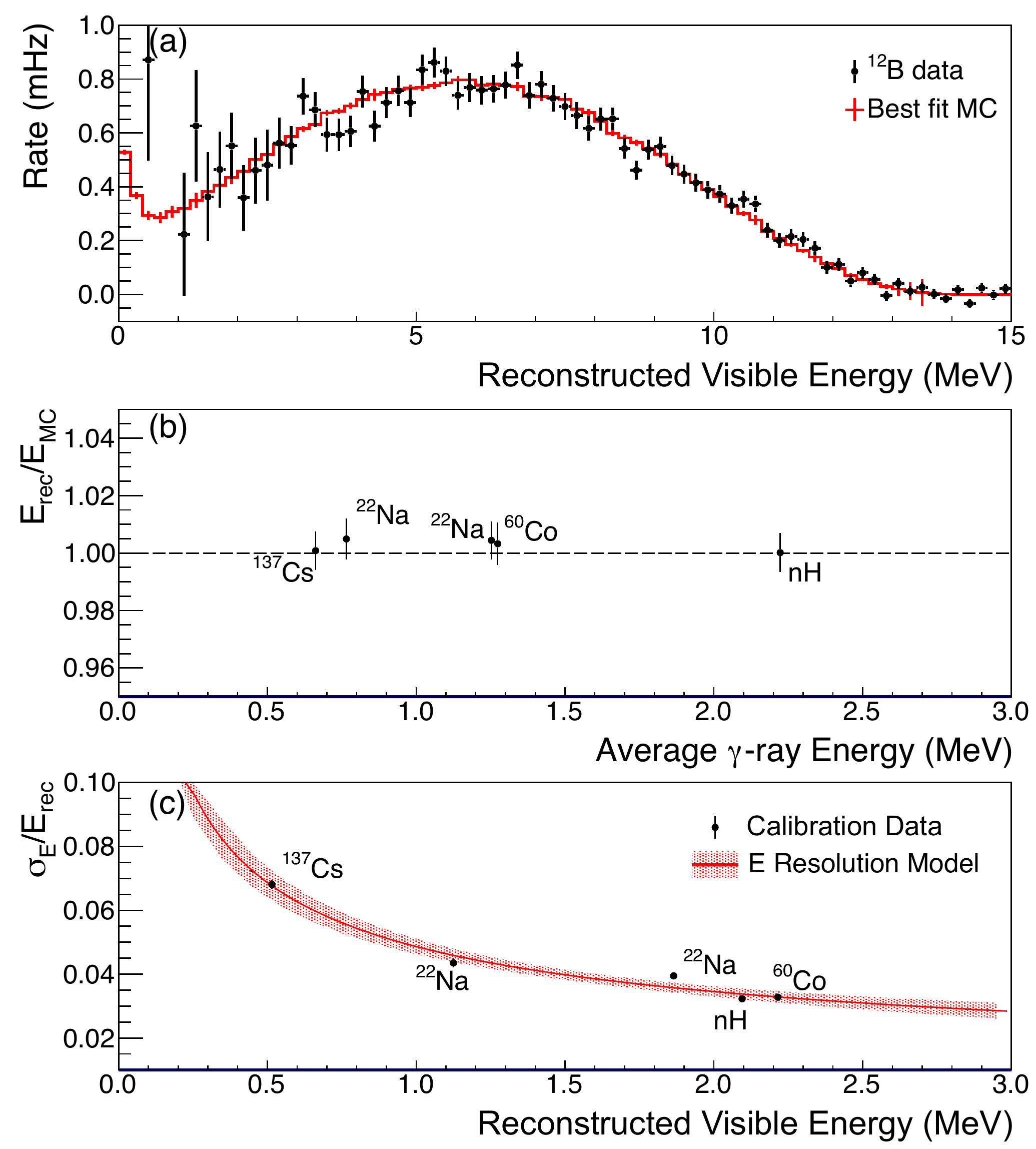}
    \caption{(a): $^{12}$B electron \erec{} spectrum compared to the best-fit MC model. (b): Ratio of measured to MC \erec{} versus average $\gamma$-ray energy, showing $\pm 1\%$ residual difference. (c): Energy resolution of the full detector versus \erec.}
    \label{fig:cal}
\end{figure}
\begin{figure}
    \centering
    \includegraphics[width=.45\textwidth]{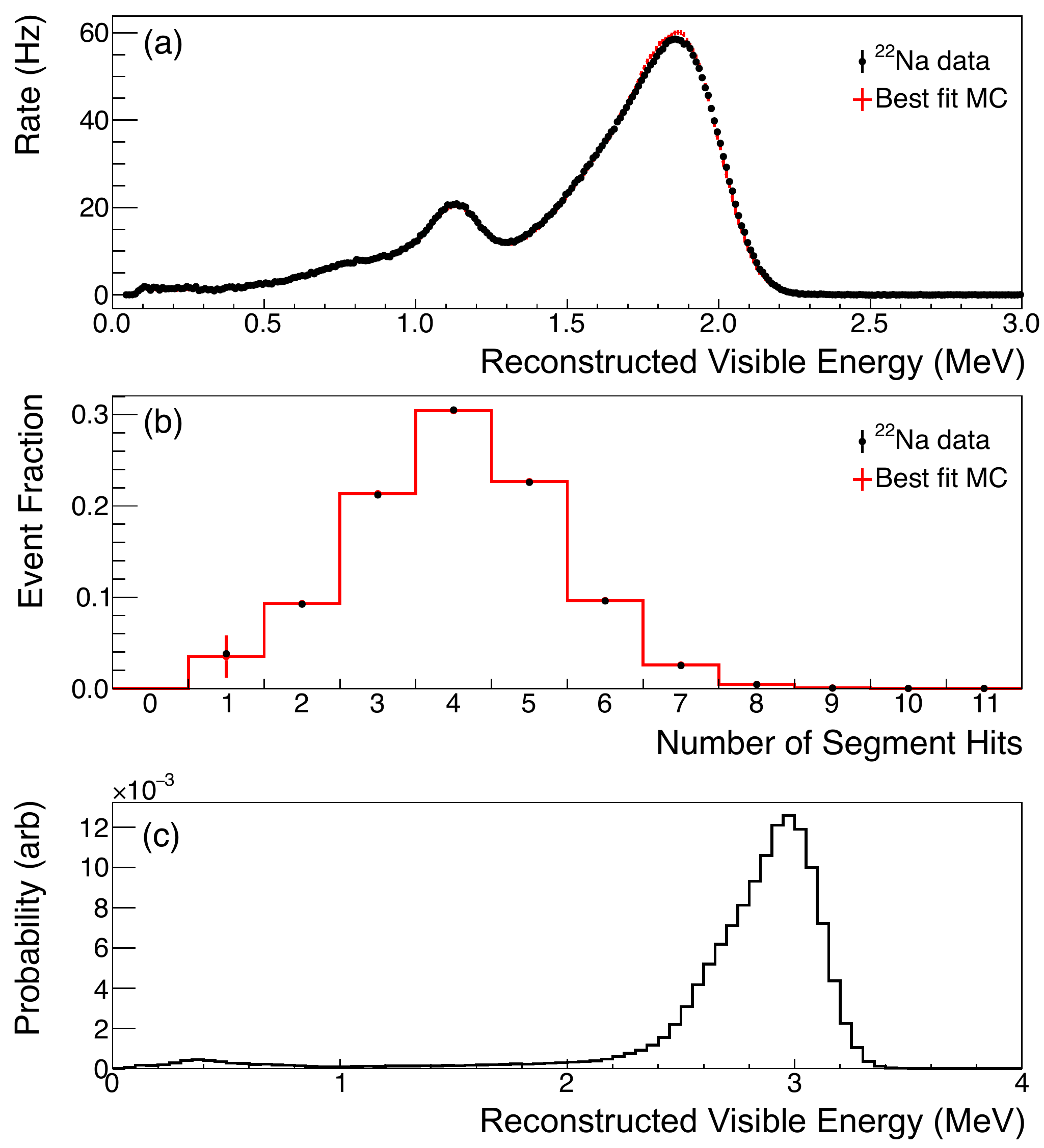}
    \caption{(a): The measured and best-fit MC simulated $^{22}$Na \erec{} spectra. (b) Distribution of segment multiplicity for simulated and measured $^{22}$Na data. (c) The simulated \erec{} of a narrow band of (4.0, 4.05)~MeV \nuebar{} energies. The distribution is shifted downwards due to scintillator non-linearity and asymmetrically broadened by escaping annihilation $\gamma$-rays.}
    \label{fig:detector_response}
\end{figure}

The measured response from deployed and environmental radioactive calibration sources are used to constrain the PROSPECT GEANT4-based~\cite{Allison:2016lfl} Monte Carlo model (PG4).
The spectra from cosmogenic $^{12}$B electrons (between [3, 13.4]\,MeV), neutron-hydrogen capture $\gamma$-rays, and centrally-deployed $\gamma$-ray sources ($^{137}$Cs, $^{60}$Co, and $^{22}$Na) are simultaneously fit to the PG4 detector response to determine the \erec{} scale, nonlinearity, and resolution summed over all detector segments (Fig.~\ref{fig:cal}). 
The segment multiplicity distributions from each calibration source are used as inputs to the fitting procedure.
The event topology of $^{22}$Na events closely resembles IBD positrons (a primary interaction accompanied by annihilation $\gamma$-rays).
Fig.~\ref{fig:detector_response} shows the comparison of the event multiplicity with the best-fit MC model.
Nonlinear scintillator response at low energies is parameterized using a combined Birks and Cherenkov model~\cite{Birks:1964zz,patrignani2016passage}, and a photo-statistics dominated energy resolution of 4.8\,\% is observed at 1\,MeV. 

As described in~\cite{Ashenfelter:2018iov}, a wide range of observables are used to track detector response stability and uniformity. 
For the data taking period considered here, \erec{} and energy resolutions are stable to within $\sim$1\,\% and $\sim$10\,\%, respectively, and z-position and z-resolution are stable to 5\,cm and $\sim$10\,\%, over all times and segments.
Small variation of segment-level resolutions are unified through the addition of event-level smearing resulting in a 5\,\% energy resolution at 1\,MeV for all segments.

A detector response matrix is constructed by generating narrow bands, 50~keV wide, of \nuebar{} energies spanning 1.8 to 10~MeV in PG4 and recording their separate \erec{} spectra.
One of these simulated \erec{} spectra (4.0-4.05~MeV \nuebar{}) is shown in Fig.~\ref{fig:detector_response}.
The main peak is shifted downward due to scintillator non-linearity and broadened from escaping annihilation $\gamma$-rays.
Events with low prompt energy ($\sim$0.5\,MeV) are observed from IBD interactions which originate in inactive material but whose annihilation $\gamma$-rays and neutrons are detected in the active volume.
The response matrix is used to convert from \nuebar{} energy to the experimental prompt \erec{} space~\cite{supplemental}.

During the data collection period considered here, a number of PMTs displayed current instabilities.
These were powered down and a total of 33 segments are excluded from the analysis. 
This has two main impacts on the analysis. 
Disabled segments reduce the efficiency of neighboring segments by lowering the acceptance of neutron captures outside of the primary interaction segment.
The increase in inactive material in the fiducial volume leads to an enhanced number of events with degraded energy, from missing either positron or annihilation $\gamma$-ray energy depositions.
The disabled segments are included in PG4 to ensure that the detector response matrix accurately captures this effect.


The selection criteria to identify IBD candidates based on the prompt positron signal and a time- and position-correlated delayed nLi signal are similar to that described in  Ref.~\cite{Ashenfelter:2018iov}. 
PSD cuts are based on measured segment-wise performance to minimize bias between segments.
A time-separation selection of (1,120)~$\micro$s and position separation cut of $\sim$15\,cm reduce accidental coincidences.
Prompt and delayed events with reconstructed positions in an outermost veto layer (1 segment width top and sides, two segment width bottom, $>$44.8\,cm from segment center) are rejected to reduce backgrounds from cosmogenic showers and $\gamma$-rays from nearby experimental activity.
IBD candidates that occur within 200$\,\micro$s after a muon interaction are vetoed to suppress multiple neutron capture events.
Candidates that occur within a (-250, 250)\,$\micro$s window of a neutron capture or nuclear recoil with \erec{} $>$ 0.25\,MeV are rejected.
The frequency of the veto conditions above is used to determine the veto dead time throughout detector operation, which ranges from 11\,\% to 14\,\% due to time-varying $\gamma$-ray backgrounds that enter the nuclear recoil band.

\begin{figure}
    \centering
    \includegraphics[width=.45\textwidth]{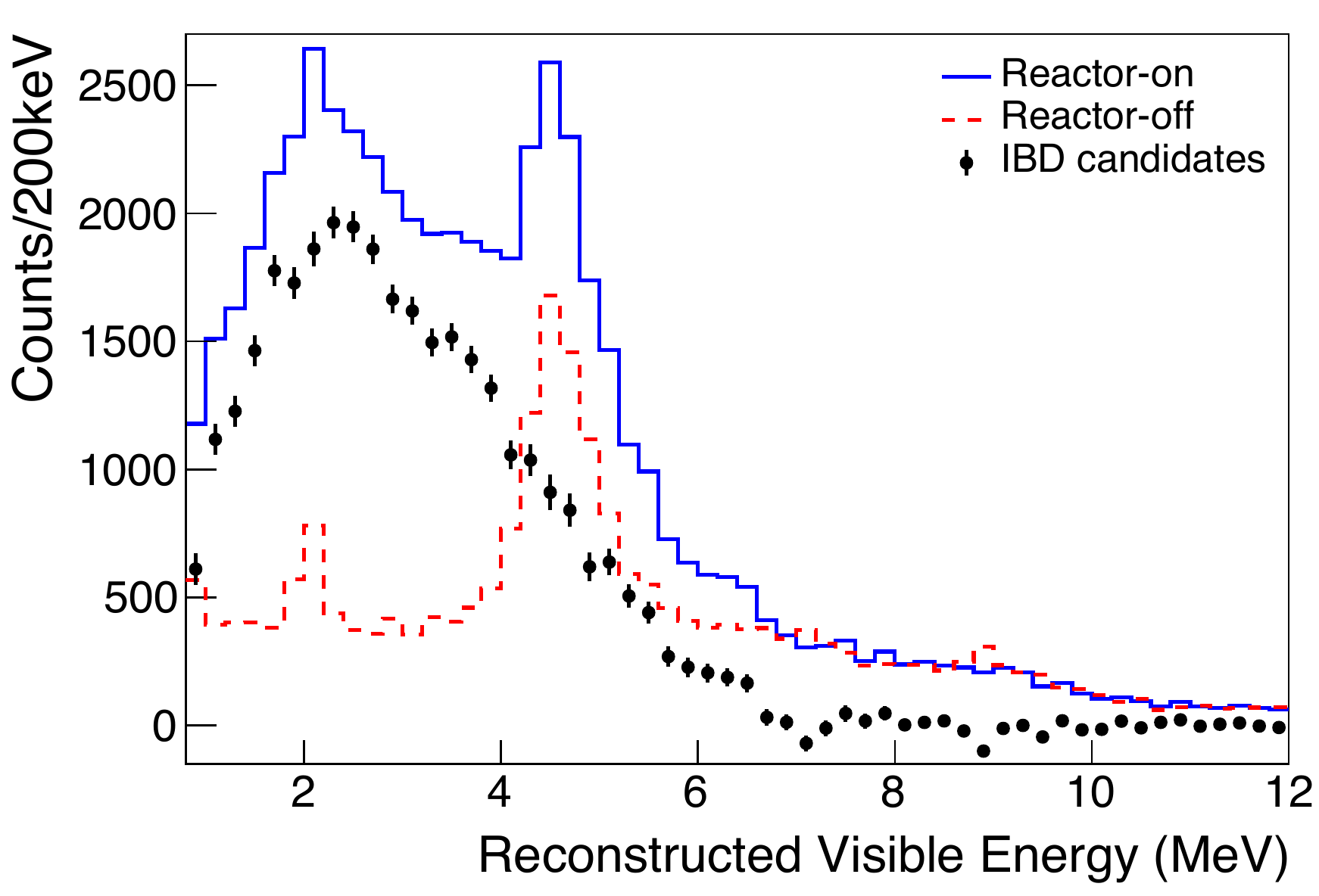}
    \caption{The reconstructed visible prompt \erec{} spectrum of IBD events (statistical errors only) compared to reactor-on and reactor-off correlated candidates. The reactor-off correlated candidates have been scaled to match the reactor-on exposure.}
    \label{fig:detected_spectrum}
\end{figure}

IBD candidate events with prompt \erec{} from 0.8 to 7.2~MeV are considered.
The 40.3 (37.8) exposure-day reactor-on (reactor-off) data set includes 70811\,$\pm$\,267\,(stat.) (20036\,$\pm$\,145\,(stat.)) IBD candidates and 20534\,$\pm$\,16\,(stat.) (1436\,$\pm$\,4\,(stat.)) accidental coincidences measured by a 10~ms wide off-time window, resulting in 50277\,$\pm$\,267\,(stat.) (18600\,$\pm$\,145\,(stat.)) correlated events.
A bin-wise subtraction of reactor-on and reactor-off correlated candidates yields the prompt \erec{} spectrum of 31678\,$\pm$\,304\,(stat.) detected \nuebar{} shown in Fig.~\ref{fig:detected_spectrum}. 
A correlated signal-to-background ratio of 1.7:1 is observed.

After all cuts are applied, the dominant backgrounds are produced by cosmogenic activity.
The IBD-like background spectrum is comprised of three components.
The 4.4~MeV peak and continuum are from fast neutron primary inelastic scattering on carbon and other material in the detector volume.
The observed 2.2~MeV peak is from multiple neutron events where the first captures on hydrogen. 
These are produced by a combination of muon and fast neutron primaries. 
The correlation of background rate with local atmospheric pressure is characterized during reactor-off periods and results in a correction factor of 0.991\,$\pm$\,0.004 based on the average pressure during reactor-on periods~\cite{Lockwood:1956, ornl_weather}.

\begin{figure}
    \centering
    \includegraphics[width=0.45\textwidth]{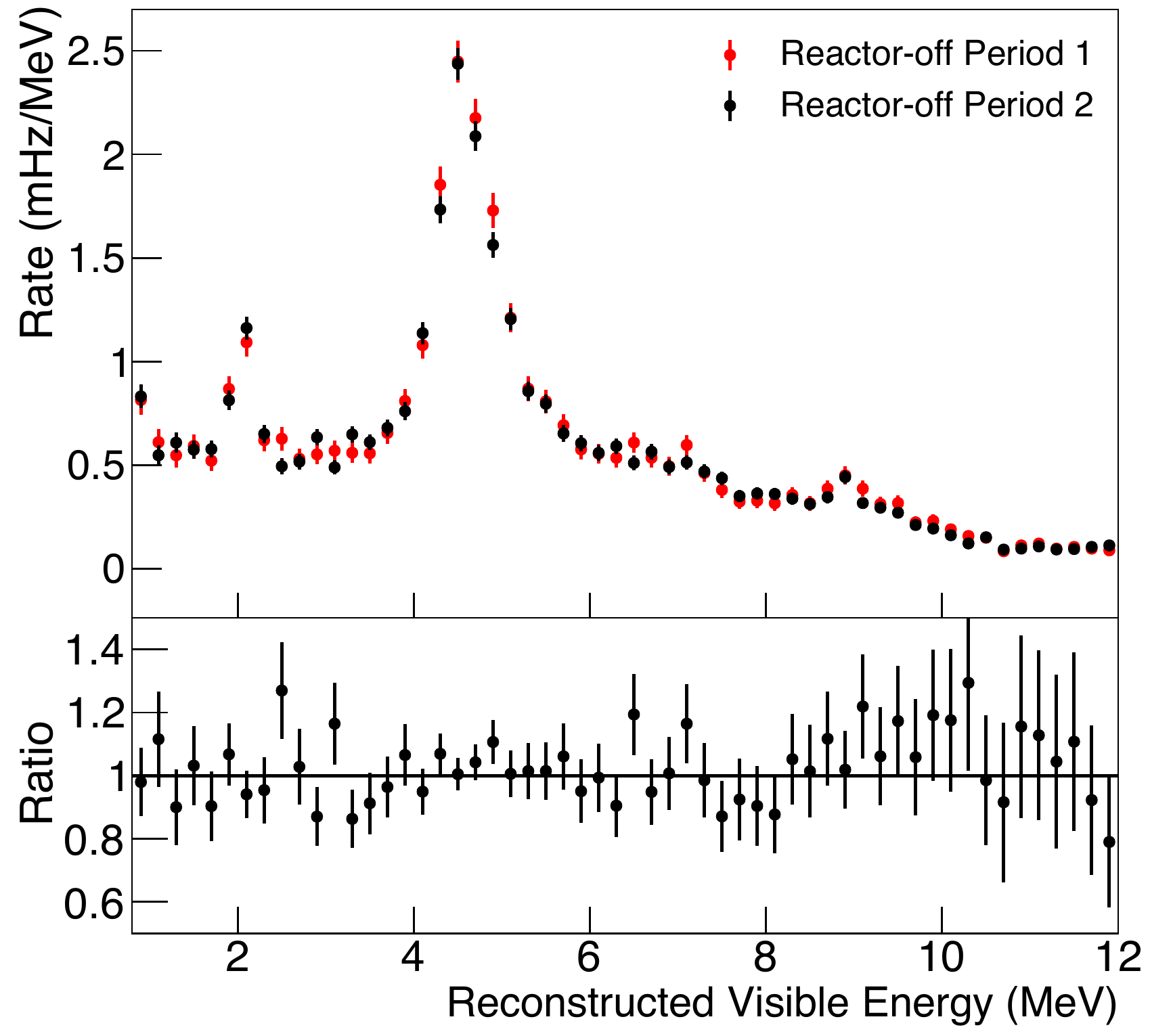}
    \caption{The reactor-off data is split into two time periods. (a) Spectra of IBD candidate events (with statistical errors only). (b) Ratio between periods. The observed consistency between these periods demonstrates the stability of cosmogenic IBD candidates after accounting for atmospheric conditions.}
    \label{fig:reactor_off_stability}
\end{figure}

Multiple validations of the energy reconstruction, background subtraction methods, and the PG4 detector model have been performed. 
This is particularly important given the segmented nature of the PROSPECT detector, the presence of inactive volume, and the prominent features present in the near-surface background spectrum.
Subdivisions of the dataset based on acquisition time and event position are used as the primary validation method.
The reactor-off data are split into two time periods (Fig.~\ref{fig:reactor_off_stability}) and their ratio between 0.8-12~MeV is compared to unity, yielding a $\chi^2/ndf$ of 35.6/56 and validating the atmospheric pressure scaling and energy reconstruction stability.
Similarly, the reactor-on and off data are each split into two independent data sets and the full analysis is performed on each portion separately.
Their ratio is compared to unity in the analysis window yielding a $\chi^2/ndf$ of 18.6/32. 
Several division schemes based on event position were also examined by splitting the detector in quadrants, near and far halves from the reactor, and inner and outer segment regions.
Consistency was found between the spectra independently measured in the first two cases, while differences in relative spectral shape due to greater energy leakage in the outer segments were successfully reproduced by the PG4 model.

A $\chi^2$ test is employed to quantify the comparison of the background-subtracted experimental data to model prediction:

\begin{align}
\label{eqn:min_chi2}
\chi^2_{min} = \Delta^{T} {V}^{-1} \Delta \\\nonumber 
\Delta_i \equiv N_i^{obs} - N_i^{pred}\times (1 + \eta) 
\end{align}

\noindent where $\Delta_i$ is the difference between the measured and predicted spectra including a free-floating nuisance parameter for normalization ($\eta$) and $V$ is the full covariance matrix. 
Simulations are performed separately varying detector parameters, including energy scale, inactive material, energy thresholds, and fiducialization according to experimental uncertainties.
Covariance matrices are generated for each parameter to capture both correlated and uncorrelated uncertainties.
The observed variation between the two reactor-off periods is used to construct a background subtraction covariance matrix.
An additional background subtraction uncertainty (4\,\%) is included to account for observed variation between the muon-induced and fast-neutron induced background components.
Finally, the uncorrelated statistical uncertainties from the reactor-on and off periods are used to build the full covariance matrix~\cite{supplemental}.

The Huber \uFive{} \nuebar{} spectrum~\cite{Huber:2011wv} is adjusted for the IBD cross-section and passed through the PROSPECT detector response matrix to response matrix to translate \nuebar{} energy to a prompt energy prediction. 
The three-neutrino framework is assumed and no correction is made for possible spectral distortions from sterile neutrino oscillations.
Corrections for non-equilibrium isotopes produced during the 24-day reactor cycle are calculated according to the procedure in Ref.~\cite{Mueller:2011nm}. 
A detailed SCALE (ORIGEN) model of the core is used to estimate the \nuebar{} flux generated via beta decay of $^{28}$Al and $^6$He nuclei produced by neutron capture on the fuel cladding and beryllium reflector~\cite{osti_1185903, SCALE61, MCNP5}.
The \nuebar{} spectra are generated according to ENDF and ENSDF data using the Oklo toolkit~\cite{Schmidt:1982zz, oklo} and added to the model with 100\,\% uncertainties.
Integrated over the full spectrum, each correction amounts to less than 1\,\% of the total \nuebar{} flux~\cite{supplemental}.
Spent nuclear fuel does not contribute a significant number of \nuebar{} interactions.

\begin{figure}[h!]
    \centering
    \includegraphics[width=.45\textwidth]{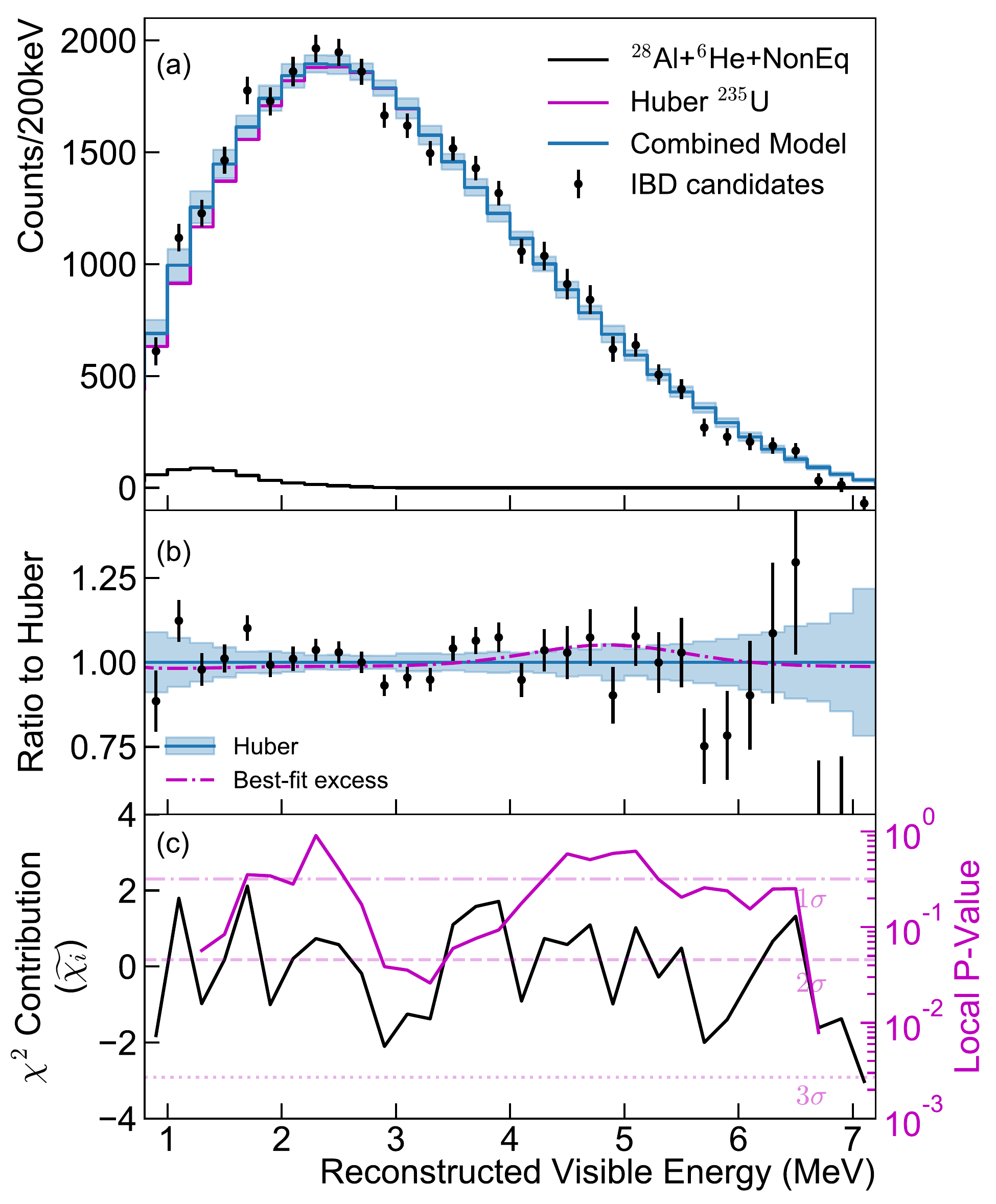}
    \caption{(a): The measured prompt energy spectrum of inverse beta decay events compared to prediction based on the Huber \uFive{}  model combined with contributions from $^{28}$Al, $^6$He, and non-equilibrium isotopes in the core. The error bars include only statistical uncertainties, while the shaded band includes detector and model uncertainties. (b): Ratio to the Huber model of the measured data and the best-fit distortion representing the spectral discrepancy observed by experiments at LEU reactors. (c): The $\chi^2$ contribution from each bin and the local p-value of a 1~MeV-wide sliding energy window.}
    \label{fig:model_compare}
\end{figure}

The shape-only comparison between the measured and predicted spectra is shown in Fig.~\ref{fig:model_compare}.
A high $\chi^2/ndf$ is observed (51.4/31), with a one-sided p-value of 0.01.
The bottom panel of Fig.~\ref{fig:model_compare} shows the signed $\chi^2$ contribution per bin ($\widetilde{\chi_i}$).
Due to non-zero off-diagonal covariance matrix elements, $\widetilde{\chi_i}$ is determined by adding an additional free-floating nuisance parameter to each bin separately and observing the change in the minimized $\chi^2$:

\begin{align}
\label{eqn:chi_per_bin}
\widetilde{\chi_i} = \frac{N_i^{obs}-N_i^{pred}}{|N_i^{obs}-N_i^{pred}|} \sqrt{\chi^2_{original} - \chi^2_{i,new}}.
\end{align}
 
To quantify the significance of local deviations from the Huber prediction, a procedure similar to Ref.~\cite{An:2016srz} is employed.
Additional free-floating nuisance parameters are included to modify $N^{pred}$ from Eqn.~\ref{eqn:min_chi2} for each bin within a 1~MeV-wide sliding energy window and a new minimum $\chi^2$ is determined. 
The change in $\chi^2$, representative of the fraction of the global $\chi^2$ contributed by that energy window, is then converted into a local p-value with five degrees of freedom, one for each bin in the window.
As shown in Fig.~\ref{fig:model_compare}, there is not one region that dominates the disagreement. 
Two regions of the spectrum have local p-values between 2-3$\sigma$, 2.8-3.5~MeV and $>$6.5~MeV. 

Spectral measurements at LEU reactors, with 50-60\,\% \uFive{} fission-fraction, have observed large deviations from predictions between 5 and 7~MeV \nuebar{} energy region. 
To compare PROSPECT's measured HEU spectrum with those from LEU reactors, an ad-hoc model is constructed from the addition of a Gaussian to the Huber \uFive{} model whose mean and width are fitted to the reported spectrum in Ref.~\cite{An:2016srz}, following studies detailed in Refs.~\cite{Buck:2015clx,Huber:2016xis}. 
The normalization of this Gaussian, $n$, is fit to the prompt \erec{} spectrum through a $\chi^2$ regression utilizing the full covariance matrix. 
A best-fit distortion, shown in Fig.~\ref{fig:model_compare}b, of $n$\,=\,0.69\,$\pm$\,0.53 is observed.
PROSPECT's behavior in this region is compatible with both the Huber \uFive{} model and a local deviation of equal size to that observed by Daya Bay ($n$\,=\,1).
A Gaussian normalization of $n$=1.78 would be required for \uFive{} to be solely responsible for the observed LEU spectral distortion, which is disfavored at 2.1$\sigma$. 
  
With a surface-based, segmented detector, PROSPECT has produced the highest statistics measurement of \uFive{} \nuebar{} spectrum to date. 
Despite broad agreement, the Huber \uFive{} model exhibits a large $\chi^2/ndf$ with respect to the measured spectrum. 
This observed \uFive{} spectrum is consistent with an ad-hoc model representing the local deviation relative to prediction observed between 5-7~MeV E$_{\nu}$ at LEU reactors.
This is a statistics-limited measurement and is expected to improve as more data are collected.

This material is based upon work supported by the following sources: US Department of Energy (DOE) Office of Science, Office of High Energy Physics under Award No. DE-SC0016357 and DE-SC0017660 to Yale University, under Award No. DE-SC0017815 to Drexel University, under Award No. DE-SC0008347 to Illinois Institute of Technology, under Award No. DE-SC0016060 to Temple University, under Contract No. DE-SC0012704 to Brookhaven National Laboratory, and under Work Proposal Number  SCW1504 to Lawrence Livermore National Laboratory. 
This work was performed under the auspices of the U.S. Department of Energy by Lawrence Livermore National Laboratory under Contract DE-AC52-07NA27344 and by Oak Ridge National Laboratory under Contract DE-AC05-00OR22725. 
Additional funding for the experiment was provided by the Heising-Simons Foundation under Award No. \#2016-117 to Yale University. 

J.G. is supported through the NSF Graduate Research Fellowship Program and A.C. performed work under appointment to the Nuclear Nonproliferation International Safeguards Fellowship Program sponsored by the National Nuclear Security Administration’s Office of International Nuclear Safeguards (NA-241). 
This work was also supported by the Canada  First  Research  Excellence  Fund  (CFREF), and the Natural Sciences and Engineering Research Council of Canada (NSERC) Discovery  program under grant \#RGPIN-418579, and Province of Ontario.

We further acknowledge support from Yale University, the Illinois Institute of Technology, Temple University, Brookhaven National Laboratory, the Lawrence Livermore National Laboratory LDRD program, the National Institute of Standards and Technology, and Oak Ridge National Laboratory.
We gratefully acknowledge the support and hospitality of the High Flux Isotope Reactor and Oak Ridge National Laboratory, managed by UT-Battelle for the U.S. Department of Energy.

\bibliographystyle{apsrev4-1}
\bibliography{refs}{}

\begin{thebibliography}{49}%
\makeatletter
\providecommand \@ifxundefined [1]{%
 \@ifx{#1\undefined}
}%
\providecommand \@ifnum [1]{%
 \ifnum #1\expandafter \@firstoftwo
 \else \expandafter \@secondoftwo
 \fi
}%
\providecommand \@ifx [1]{%
 \ifx #1\expandafter \@firstoftwo
 \else \expandafter \@secondoftwo
 \fi
}%
\providecommand \natexlab [1]{#1}%
\providecommand \enquote  [1]{``#1''}%
\providecommand \bibnamefont  [1]{#1}%
\providecommand \bibfnamefont [1]{#1}%
\providecommand \citenamefont [1]{#1}%
\providecommand \href@noop [0]{\@secondoftwo}%
\providecommand \href [0]{\begingroup \@sanitize@url \@href}%
\providecommand \@href[1]{\@@startlink{#1}\@@href}%
\providecommand \@@href[1]{\endgroup#1\@@endlink}%
\providecommand \@sanitize@url [0]{\catcode `\\12\catcode `\$12\catcode
  `\&12\catcode `\#12\catcode `\^12\catcode `\_12\catcode `\%12\relax}%
\providecommand \@@startlink[1]{}%
\providecommand \@@endlink[0]{}%
\providecommand \url  [0]{\begingroup\@sanitize@url \@url }%
\providecommand \@url [1]{\endgroup\@href {#1}{\urlprefix }}%
\providecommand \urlprefix  [0]{URL }%
\providecommand \Eprint [0]{\href }%
\providecommand \doibase [0]{http://dx.doi.org/}%
\providecommand \selectlanguage [0]{\@gobble}%
\providecommand \bibinfo  [0]{\@secondoftwo}%
\providecommand \bibfield  [0]{\@secondoftwo}%
\providecommand \translation [1]{[#1]}%
\providecommand \BibitemOpen [0]{}%
\providecommand \bibitemStop [0]{}%
\providecommand \bibitemNoStop [0]{.\EOS\space}%
\providecommand \EOS [0]{\spacefactor3000\relax}%
\providecommand \BibitemShut  [1]{\csname bibitem#1\endcsname}%
\let\auto@bib@innerbib\@empty
\bibitem [{\citenamefont {Cowan}\ \emph {et~al.}(1956)\citenamefont {Cowan},
  \citenamefont {Reines}, \citenamefont {Harrison}, \citenamefont {Kruse},\
  and\ \citenamefont {McGuire}}]{Cowan:1992xc}%
  \BibitemOpen
  \bibfield  {author} {\bibinfo {author} {\bibfnamefont {C.~L.}\ \bibnamefont
  {Cowan}}, \bibinfo {author} {\bibfnamefont {F.}~\bibnamefont {Reines}},
  \bibinfo {author} {\bibfnamefont {F.~B.}\ \bibnamefont {Harrison}}, \bibinfo
  {author} {\bibfnamefont {H.~W.}\ \bibnamefont {Kruse}}, \ and\ \bibinfo
  {author} {\bibfnamefont {A.~D.}\ \bibnamefont {McGuire}},\ }\href {\doibase
  10.1126/science.124.3212.103} {\bibfield  {journal} {\bibinfo  {journal}
  {Science}\ }\textbf {\bibinfo {volume} {124}},\ \bibinfo {pages} {103}
  (\bibinfo {year} {1956})}\BibitemShut {NoStop}%
\bibitem [{\citenamefont {Eguchi}\ \emph {et~al.}(2003)\citenamefont {Eguchi}
  \emph {et~al.}}]{Eguchi:2002dm}%
  \BibitemOpen
  \bibfield  {author} {\bibinfo {author} {\bibfnamefont {K.}~\bibnamefont
  {Eguchi}} \emph {et~al.} (\bibinfo {collaboration} {KamLAND}),\ }\href
  {\doibase 10.1103/PhysRevLett.90.021802} {\bibfield  {journal} {\bibinfo
  {journal} {Phys. Rev. Lett.}\ }\textbf {\bibinfo {volume} {90}},\ \bibinfo
  {pages} {021802} (\bibinfo {year} {2003})},\ \Eprint
  {http://arxiv.org/abs/hep-ex/0212021} {arXiv:hep-ex/0212021 [hep-ex]}
  \BibitemShut {NoStop}%
\bibitem [{\citenamefont {Araki}\ \emph {et~al.}(2005)\citenamefont {Araki}
  \emph {et~al.}}]{Araki:2005qa}%
  \BibitemOpen
  \bibfield  {author} {\bibinfo {author} {\bibfnamefont {T.}~\bibnamefont
  {Araki}} \emph {et~al.},\ }\href {\doibase 10.1038/nature03980} {\bibfield
  {journal} {\bibinfo  {journal} {Nature}\ }\textbf {\bibinfo {volume} {436}},\
  \bibinfo {pages} {499} (\bibinfo {year} {2005})}\BibitemShut {NoStop}%
\bibitem [{\citenamefont {An}\ \emph {et~al.}(2012)\citenamefont {An} \emph
  {et~al.}}]{An:2012eh}%
  \BibitemOpen
  \bibfield  {author} {\bibinfo {author} {\bibfnamefont {F.~P.}\ \bibnamefont
  {An}} \emph {et~al.} (\bibinfo {collaboration} {Daya Bay}),\ }\href {\doibase
  10.1103/PhysRevLett.108.171803} {\bibfield  {journal} {\bibinfo  {journal}
  {Phys. Rev. Lett.}\ }\textbf {\bibinfo {volume} {108}},\ \bibinfo {pages}
  {171803} (\bibinfo {year} {2012})},\ \Eprint {http://arxiv.org/abs/1203.1669}
  {arXiv:1203.1669 [hep-ex]} \BibitemShut {NoStop}%
\bibitem [{\citenamefont {Ahn}\ \emph {et~al.}(2012)\citenamefont {Ahn} \emph
  {et~al.}}]{Ahn:2012nd}%
  \BibitemOpen
  \bibfield  {author} {\bibinfo {author} {\bibfnamefont {J.~K.}\ \bibnamefont
  {Ahn}} \emph {et~al.} (\bibinfo {collaboration} {RENO}),\ }\href {\doibase
  10.1103/PhysRevLett.108.191802} {\bibfield  {journal} {\bibinfo  {journal}
  {Phys. Rev. Lett.}\ }\textbf {\bibinfo {volume} {108}},\ \bibinfo {pages}
  {191802} (\bibinfo {year} {2012})},\ \Eprint {http://arxiv.org/abs/1204.0626}
  {arXiv:1204.0626 [hep-ex]} \BibitemShut {NoStop}%
\bibitem [{\citenamefont {Abe}\ \emph {et~al.}(2012)\citenamefont {Abe} \emph
  {et~al.}}]{Abe:2011fz}%
  \BibitemOpen
  \bibfield  {author} {\bibinfo {author} {\bibfnamefont {Y.}~\bibnamefont
  {Abe}} \emph {et~al.} (\bibinfo {collaboration} {Double Chooz}),\ }\href
  {\doibase 10.1103/PhysRevLett.108.131801} {\bibfield  {journal} {\bibinfo
  {journal} {Phys. Rev. Lett.}\ }\textbf {\bibinfo {volume} {108}},\ \bibinfo
  {pages} {131801} (\bibinfo {year} {2012})},\ \Eprint
  {http://arxiv.org/abs/1112.6353} {arXiv:1112.6353 [hep-ex]} \BibitemShut
  {NoStop}%
\bibitem [{\citenamefont {Mueller}\ \emph {et~al.}(2011)\citenamefont {Mueller}
  \emph {et~al.}}]{Mueller:2011nm}%
  \BibitemOpen
  \bibfield  {author} {\bibinfo {author} {\bibfnamefont {T.~A.}\ \bibnamefont
  {Mueller}} \emph {et~al.},\ }\href {\doibase 10.1103/PhysRevC.83.054615}
  {\bibfield  {journal} {\bibinfo  {journal} {Phys. Rev.}\ }\textbf {\bibinfo
  {volume} {C83}},\ \bibinfo {pages} {054615} (\bibinfo {year} {2011})},\
  \Eprint {http://arxiv.org/abs/1101.2663} {arXiv:1101.2663 [hep-ex]}
  \BibitemShut {NoStop}%
\bibitem [{\citenamefont {Mention}\ \emph {et~al.}(2011)\citenamefont
  {Mention}, \citenamefont {Fechner}, \citenamefont {Lasserre}, \citenamefont
  {Mueller}, \citenamefont {Lhuillier}, \citenamefont {Cribier},\ and\
  \citenamefont {Letourneau}}]{Mention:2011rk}%
  \BibitemOpen
  \bibfield  {author} {\bibinfo {author} {\bibfnamefont {G.}~\bibnamefont
  {Mention}}, \bibinfo {author} {\bibfnamefont {M.}~\bibnamefont {Fechner}},
  \bibinfo {author} {\bibfnamefont {T.}~\bibnamefont {Lasserre}}, \bibinfo
  {author} {\bibfnamefont {T.~A.}\ \bibnamefont {Mueller}}, \bibinfo {author}
  {\bibfnamefont {D.}~\bibnamefont {Lhuillier}}, \bibinfo {author}
  {\bibfnamefont {M.}~\bibnamefont {Cribier}}, \ and\ \bibinfo {author}
  {\bibfnamefont {A.}~\bibnamefont {Letourneau}},\ }\href {\doibase
  10.1103/PhysRevD.83.073006} {\bibfield  {journal} {\bibinfo  {journal} {Phys.
  Rev.}\ }\textbf {\bibinfo {volume} {D83}},\ \bibinfo {pages} {073006}
  (\bibinfo {year} {2011})},\ \Eprint {http://arxiv.org/abs/1101.2755}
  {arXiv:1101.2755 [hep-ex]} \BibitemShut {NoStop}%
\bibitem [{\citenamefont {Huber}(2011)}]{Huber:2011wv}%
  \BibitemOpen
  \bibfield  {author} {\bibinfo {author} {\bibfnamefont {P.}~\bibnamefont
  {Huber}},\ }\href {\doibase 10.1103/PhysRevC.85.029901,
  10.1103/PhysRevC.84.024617} {\bibfield  {journal} {\bibinfo  {journal} {Phys.
  Rev.}\ }\textbf {\bibinfo {volume} {C84}},\ \bibinfo {pages} {024617}
  (\bibinfo {year} {2011})},\ \bibinfo {note} {[Erratum: Phys.
  Rev.C85,029901(2012)]},\ \Eprint {http://arxiv.org/abs/1106.0687}
  {arXiv:1106.0687 [hep-ph]} \BibitemShut {NoStop}%
\bibitem [{\citenamefont {Fallot}\ \emph {et~al.}(2012)\citenamefont {Fallot}
  \emph {et~al.}}]{Fallot:2012jv}%
  \BibitemOpen
  \bibfield  {author} {\bibinfo {author} {\bibfnamefont {M.}~\bibnamefont
  {Fallot}} \emph {et~al.},\ }\href {\doibase 10.1103/PhysRevLett.109.202504}
  {\bibfield  {journal} {\bibinfo  {journal} {Phys. Rev. Lett.}\ }\textbf
  {\bibinfo {volume} {109}},\ \bibinfo {pages} {202504} (\bibinfo {year}
  {2012})},\ \Eprint {http://arxiv.org/abs/1208.3877} {arXiv:1208.3877
  [nucl-ex]} \BibitemShut {NoStop}%
\bibitem [{\citenamefont {An}\ \emph {et~al.}(2016)\citenamefont {An} \emph
  {et~al.}}]{An:2015nua}%
  \BibitemOpen
  \bibfield  {author} {\bibinfo {author} {\bibfnamefont {F.~P.}\ \bibnamefont
  {An}} \emph {et~al.} (\bibinfo {collaboration} {Daya Bay}),\ }\href {\doibase
  10.1103/PhysRevLett.116.061801, 10.1103/PhysRevLett.118.099902} {\bibfield
  {journal} {\bibinfo  {journal} {Phys. Rev. Lett.}\ }\textbf {\bibinfo
  {volume} {116}},\ \bibinfo {pages} {061801} (\bibinfo {year} {2016})},\
  \bibinfo {note} {[Erratum: Phys. Rev. Lett.118,no.9,099902(2017)]},\ \Eprint
  {http://arxiv.org/abs/1508.04233} {arXiv:1508.04233 [hep-ex]} \BibitemShut
  {NoStop}%
\bibitem [{\citenamefont {Abe}\ \emph {et~al.}(2014)\citenamefont {Abe} \emph
  {et~al.}}]{Abe:2014bwa}%
  \BibitemOpen
  \bibfield  {author} {\bibinfo {author} {\bibfnamefont {Y.}~\bibnamefont
  {Abe}} \emph {et~al.} (\bibinfo {collaboration} {Double Chooz}),\ }\href
  {\doibase 10.1007/JHEP02(2015)074, 10.1007/JHEP10(2014)086} {\bibfield
  {journal} {\bibinfo  {journal} {JHEP}\ }\textbf {\bibinfo {volume} {10}},\
  \bibinfo {pages} {086} (\bibinfo {year} {2014})},\ \bibinfo {note} {[Erratum:
  JHEP02,074(2015)]},\ \Eprint {http://arxiv.org/abs/1406.7763}
  {arXiv:1406.7763 [hep-ex]} \BibitemShut {NoStop}%
\bibitem [{\citenamefont {Seo}(2015)}]{Seo:2014xei}%
  \BibitemOpen
  \bibfield  {author} {\bibinfo {author} {\bibfnamefont {S.-H.}\ \bibnamefont
  {Seo}} (\bibinfo {collaboration} {RENO}),\ }\bibfield  {booktitle} {\emph
  {\bibinfo {booktitle} {{Proceedings, 26th International Conference on
  Neutrino Physics and Astrophysics (Neutrino 2014): Boston, Massachusetts,
  United States, June 2-7, 2014}}},\ }\href {\doibase 10.1063/1.4915563}
  {\bibfield  {journal} {\bibinfo  {journal} {AIP Conf. Proc.}\ }\textbf
  {\bibinfo {volume} {1666}},\ \bibinfo {pages} {080002} (\bibinfo {year}
  {2015})},\ \Eprint {http://arxiv.org/abs/1410.7987} {arXiv:1410.7987
  [hep-ex]} \BibitemShut {NoStop}%
\bibitem [{\citenamefont {Ko}\ \emph {et~al.}(2017)\citenamefont {Ko} \emph
  {et~al.}}]{Ko:2016owz}%
  \BibitemOpen
  \bibfield  {author} {\bibinfo {author} {\bibfnamefont {Y.}~\bibnamefont {Ko}}
  \emph {et~al.} (\bibinfo {collaboration} {NEOS}),\ }\href {\doibase
  10.1103/PhysRevLett.118.121802} {\bibfield  {journal} {\bibinfo  {journal}
  {Phys. Rev. Lett.}\ }\textbf {\bibinfo {volume} {118}},\ \bibinfo {pages}
  {121802} (\bibinfo {year} {2017})},\ \Eprint
  {http://arxiv.org/abs/1610.05134} {arXiv:1610.05134 [hep-ex]} \BibitemShut
  {NoStop}%
\bibitem [{\citenamefont {An}\ \emph {et~al.}(2017{\natexlab{a}})\citenamefont
  {An} \emph {et~al.}}]{An:2017osx}%
  \BibitemOpen
  \bibfield  {author} {\bibinfo {author} {\bibfnamefont {F.~P.}\ \bibnamefont
  {An}} \emph {et~al.} (\bibinfo {collaboration} {Daya Bay}),\ }\href {\doibase
  10.1103/PhysRevLett.118.251801} {\bibfield  {journal} {\bibinfo  {journal}
  {Phys. Rev. Lett.}\ }\textbf {\bibinfo {volume} {118}},\ \bibinfo {pages}
  {251801} (\bibinfo {year} {2017}{\natexlab{a}})},\ \Eprint
  {http://arxiv.org/abs/1704.01082} {arXiv:1704.01082 [hep-ex]} \BibitemShut
  {NoStop}%
\bibitem [{\citenamefont {Bak}\ \emph {et~al.}(2018)\citenamefont {Bak} \emph
  {et~al.}}]{RENO:2018pwo}%
  \BibitemOpen
  \bibfield  {author} {\bibinfo {author} {\bibfnamefont {G.}~\bibnamefont
  {Bak}} \emph {et~al.} (\bibinfo {collaboration} {RENO}),\ }\href@noop {} {\
  (\bibinfo {year} {2018})},\ \Eprint {http://arxiv.org/abs/1806.00574}
  {arXiv:1806.00574 [hep-ex]} \BibitemShut {NoStop}%
\bibitem [{\citenamefont {Hayes}\ \emph {et~al.}(2014)\citenamefont {Hayes},
  \citenamefont {Friar}, \citenamefont {Garvey}, \citenamefont {Jungman},\ and\
  \citenamefont {Jonkmans}}]{Hayes:2013wra}%
  \BibitemOpen
  \bibfield  {author} {\bibinfo {author} {\bibfnamefont {A.~C.}\ \bibnamefont
  {Hayes}}, \bibinfo {author} {\bibfnamefont {J.~L.}\ \bibnamefont {Friar}},
  \bibinfo {author} {\bibfnamefont {G.~T.}\ \bibnamefont {Garvey}}, \bibinfo
  {author} {\bibfnamefont {G.}~\bibnamefont {Jungman}}, \ and\ \bibinfo
  {author} {\bibfnamefont {G.}~\bibnamefont {Jonkmans}},\ }\href {\doibase
  10.1103/PhysRevLett.112.202501} {\bibfield  {journal} {\bibinfo  {journal}
  {Phys. Rev. Lett.}\ }\textbf {\bibinfo {volume} {112}},\ \bibinfo {pages}
  {202501} (\bibinfo {year} {2014})},\ \Eprint {http://arxiv.org/abs/1309.4146}
  {arXiv:1309.4146 [nucl-th]} \BibitemShut {NoStop}%
\bibitem [{\citenamefont {Dwyer}\ and\ \citenamefont
  {Langford}(2015)}]{Dwyer:2014eka}%
  \BibitemOpen
  \bibfield  {author} {\bibinfo {author} {\bibfnamefont {D.~A.}\ \bibnamefont
  {Dwyer}}\ and\ \bibinfo {author} {\bibfnamefont {T.~J.}\ \bibnamefont
  {Langford}},\ }\href {\doibase 10.1103/PhysRevLett.114.012502} {\bibfield
  {journal} {\bibinfo  {journal} {Phys. Rev. Lett.}\ }\textbf {\bibinfo
  {volume} {114}},\ \bibinfo {pages} {012502} (\bibinfo {year} {2015})},\
  \Eprint {http://arxiv.org/abs/1407.1281} {arXiv:1407.1281 [nucl-ex]}
  \BibitemShut {NoStop}%
\bibitem [{\citenamefont {Hayes}\ \emph {et~al.}(2015)\citenamefont {Hayes},
  \citenamefont {Friar}, \citenamefont {Garvey}, \citenamefont {Ibeling},
  \citenamefont {Jungman}, \citenamefont {Kawano},\ and\ \citenamefont
  {Mills}}]{Hayes:2015yka}%
  \BibitemOpen
  \bibfield  {author} {\bibinfo {author} {\bibfnamefont {A.~C.}\ \bibnamefont
  {Hayes}}, \bibinfo {author} {\bibfnamefont {J.~L.}\ \bibnamefont {Friar}},
  \bibinfo {author} {\bibfnamefont {G.~T.}\ \bibnamefont {Garvey}}, \bibinfo
  {author} {\bibfnamefont {D.}~\bibnamefont {Ibeling}}, \bibinfo {author}
  {\bibfnamefont {G.}~\bibnamefont {Jungman}}, \bibinfo {author} {\bibfnamefont
  {T.}~\bibnamefont {Kawano}}, \ and\ \bibinfo {author} {\bibfnamefont {R.~W.}\
  \bibnamefont {Mills}},\ }\href {\doibase 10.1103/PhysRevD.92.033015}
  {\bibfield  {journal} {\bibinfo  {journal} {Phys. Rev.}\ }\textbf {\bibinfo
  {volume} {D92}},\ \bibinfo {pages} {033015} (\bibinfo {year} {2015})},\
  \Eprint {http://arxiv.org/abs/1506.00583} {arXiv:1506.00583 [nucl-th]}
  \BibitemShut {NoStop}%
\bibitem [{\citenamefont {Wang}\ \emph {et~al.}(2016)\citenamefont {Wang},
  \citenamefont {Friar},\ and\ \citenamefont {Hayes}}]{Wang:2016rqh}%
  \BibitemOpen
  \bibfield  {author} {\bibinfo {author} {\bibfnamefont {X.~B.}\ \bibnamefont
  {Wang}}, \bibinfo {author} {\bibfnamefont {J.~L.}\ \bibnamefont {Friar}}, \
  and\ \bibinfo {author} {\bibfnamefont {A.~C.}\ \bibnamefont {Hayes}},\ }\href
  {\doibase 10.1103/PhysRevC.94.034314} {\bibfield  {journal} {\bibinfo
  {journal} {Phys. Rev.}\ }\textbf {\bibinfo {volume} {C94}},\ \bibinfo {pages}
  {034314} (\bibinfo {year} {2016})},\ \Eprint
  {http://arxiv.org/abs/1607.02149} {arXiv:1607.02149 [nucl-th]} \BibitemShut
  {NoStop}%
\bibitem [{\citenamefont {Wang}\ and\ \citenamefont
  {Hayes}(2017)}]{Wang:2017htp}%
  \BibitemOpen
  \bibfield  {author} {\bibinfo {author} {\bibfnamefont {X.~B.}\ \bibnamefont
  {Wang}}\ and\ \bibinfo {author} {\bibfnamefont {A.~C.}\ \bibnamefont
  {Hayes}},\ }\href {\doibase 10.1103/PhysRevC.95.064313} {\bibfield  {journal}
  {\bibinfo  {journal} {Phys. Rev.}\ }\textbf {\bibinfo {volume} {C95}},\
  \bibinfo {pages} {064313} (\bibinfo {year} {2017})},\ \Eprint
  {http://arxiv.org/abs/1702.07520} {arXiv:1702.07520 [nucl-th]} \BibitemShut
  {NoStop}%
\bibitem [{\citenamefont {Sonzogni}\ \emph {et~al.}(2017)\citenamefont
  {Sonzogni}, \citenamefont {McCutchan},\ and\ \citenamefont
  {Hayes}}]{Sonzogni:2017wxy}%
  \BibitemOpen
  \bibfield  {author} {\bibinfo {author} {\bibfnamefont {A.}~\bibnamefont
  {Sonzogni}}, \bibinfo {author} {\bibfnamefont {E.}~\bibnamefont {McCutchan}},
  \ and\ \bibinfo {author} {\bibfnamefont {A.}~\bibnamefont {Hayes}},\ }\href
  {\doibase 10.1103/PhysRevLett.119.112501} {\bibfield  {journal} {\bibinfo
  {journal} {Phys. Rev. Lett.}\ }\textbf {\bibinfo {volume} {119}},\ \bibinfo
  {pages} {112501} (\bibinfo {year} {2017})}\BibitemShut {NoStop}%
\bibitem [{\citenamefont {Gebre}\ \emph {et~al.}(2018)\citenamefont {Gebre},
  \citenamefont {Littlejohn},\ and\ \citenamefont {Surukuchi}}]{Gebre:2017vmm}%
  \BibitemOpen
  \bibfield  {author} {\bibinfo {author} {\bibfnamefont {Y.}~\bibnamefont
  {Gebre}}, \bibinfo {author} {\bibfnamefont {B.~R.}\ \bibnamefont
  {Littlejohn}}, \ and\ \bibinfo {author} {\bibfnamefont {P.~T.}\ \bibnamefont
  {Surukuchi}},\ }\href {\doibase 10.1103/PhysRevD.97.013003} {\bibfield
  {journal} {\bibinfo  {journal} {Phys. Rev.}\ }\textbf {\bibinfo {volume}
  {D97}},\ \bibinfo {pages} {013003} (\bibinfo {year} {2018})},\ \Eprint
  {http://arxiv.org/abs/1709.10051} {arXiv:1709.10051 [hep-ph]} \BibitemShut
  {NoStop}%
\bibitem [{\citenamefont {Littlejohn}\ \emph {et~al.}(2018)\citenamefont
  {Littlejohn}, \citenamefont {Conant}, \citenamefont {Dwyer}, \citenamefont
  {Erickson}, \citenamefont {Gustafson},\ and\ \citenamefont
  {Hermanek}}]{Littlejohn:2018hqm}%
  \BibitemOpen
  \bibfield  {author} {\bibinfo {author} {\bibfnamefont {B.~R.}\ \bibnamefont
  {Littlejohn}}, \bibinfo {author} {\bibfnamefont {A.}~\bibnamefont {Conant}},
  \bibinfo {author} {\bibfnamefont {D.~A.}\ \bibnamefont {Dwyer}}, \bibinfo
  {author} {\bibfnamefont {A.}~\bibnamefont {Erickson}}, \bibinfo {author}
  {\bibfnamefont {I.}~\bibnamefont {Gustafson}}, \ and\ \bibinfo {author}
  {\bibfnamefont {K.}~\bibnamefont {Hermanek}},\ }\href {\doibase
  10.1103/PhysRevD.97.073007} {\bibfield  {journal} {\bibinfo  {journal} {Phys.
  Rev.}\ }\textbf {\bibinfo {volume} {D97}},\ \bibinfo {pages} {073007}
  (\bibinfo {year} {2018})},\ \Eprint {http://arxiv.org/abs/1803.01787}
  {arXiv:1803.01787 [nucl-th]} \BibitemShut {NoStop}%
\bibitem [{\citenamefont {Giunti}\ \emph {et~al.}(2017)\citenamefont {Giunti},
  \citenamefont {Ji}, \citenamefont {Laveder}, \citenamefont {Li},\ and\
  \citenamefont {Littlejohn}}]{Giunti:2017yid}%
  \BibitemOpen
  \bibfield  {author} {\bibinfo {author} {\bibfnamefont {C.}~\bibnamefont
  {Giunti}}, \bibinfo {author} {\bibfnamefont {X.~P.}\ \bibnamefont {Ji}},
  \bibinfo {author} {\bibfnamefont {M.}~\bibnamefont {Laveder}}, \bibinfo
  {author} {\bibfnamefont {Y.~F.}\ \bibnamefont {Li}}, \ and\ \bibinfo {author}
  {\bibfnamefont {B.~R.}\ \bibnamefont {Littlejohn}},\ }\href {\doibase
  10.1007/JHEP10(2017)143} {\bibfield  {journal} {\bibinfo  {journal} {JHEP}\
  }\textbf {\bibinfo {volume} {10}},\ \bibinfo {pages} {143} (\bibinfo {year}
  {2017})},\ \Eprint {http://arxiv.org/abs/1708.01133} {arXiv:1708.01133
  [hep-ph]} \BibitemShut {NoStop}%
\bibitem [{\citenamefont {Dentler}\ \emph {et~al.}(2017)\citenamefont
  {Dentler}, \citenamefont {Hernández-Cabezudo}, \citenamefont {Kopp},
  \citenamefont {Maltoni},\ and\ \citenamefont {Schwetz}}]{Dentler:2017tkw}%
  \BibitemOpen
  \bibfield  {author} {\bibinfo {author} {\bibfnamefont {M.}~\bibnamefont
  {Dentler}}, \bibinfo {author} {\bibfnamefont {A.}~\bibnamefont
  {Hernández-Cabezudo}}, \bibinfo {author} {\bibfnamefont {J.}~\bibnamefont
  {Kopp}}, \bibinfo {author} {\bibfnamefont {M.}~\bibnamefont {Maltoni}}, \
  and\ \bibinfo {author} {\bibfnamefont {T.}~\bibnamefont {Schwetz}},\ }\href
  {\doibase 10.1007/JHEP11(2017)099} {\bibfield  {journal} {\bibinfo  {journal}
  {JHEP}\ }\textbf {\bibinfo {volume} {11}},\ \bibinfo {pages} {099} (\bibinfo
  {year} {2017})},\ \Eprint {http://arxiv.org/abs/1709.04294} {arXiv:1709.04294
  [hep-ph]} \BibitemShut {NoStop}%
\bibitem [{\citenamefont {Giunti}\ \emph {et~al.}(2019)\citenamefont {Giunti},
  \citenamefont {Li}, \citenamefont {Littlejohn},\ and\ \citenamefont
  {Surukuchi}}]{Giunti:2019qlt}%
  \BibitemOpen
  \bibfield  {author} {\bibinfo {author} {\bibfnamefont {C.}~\bibnamefont
  {Giunti}}, \bibinfo {author} {\bibfnamefont {Y.~F.}\ \bibnamefont {Li}},
  \bibinfo {author} {\bibfnamefont {B.~R.}\ \bibnamefont {Littlejohn}}, \ and\
  \bibinfo {author} {\bibfnamefont {P.~T.}\ \bibnamefont {Surukuchi}},\
  }\href@noop {} {\  (\bibinfo {year} {2019})},\ \Eprint
  {http://arxiv.org/abs/1901.01807} {arXiv:1901.01807 [hep-ph]} \BibitemShut
  {NoStop}%
\bibitem [{\citenamefont {Allemandou}\ \emph {et~al.}(2018)\citenamefont
  {Allemandou} \emph {et~al.}}]{Allemandou:2018vwb}%
  \BibitemOpen
  \bibfield  {author} {\bibinfo {author} {\bibfnamefont {N.}~\bibnamefont
  {Allemandou}} \emph {et~al.} (\bibinfo {collaboration} {STEREO}),\ }\href
  {\doibase 10.1088/1748-0221/13/07/P07009} {\bibfield  {journal} {\bibinfo
  {journal} {JINST}\ }\textbf {\bibinfo {volume} {13}},\ \bibinfo {pages}
  {P07009} (\bibinfo {year} {2018})},\ \Eprint
  {http://arxiv.org/abs/1804.09052} {arXiv:1804.09052 [physics.ins-det]}
  \BibitemShut {NoStop}%
\bibitem [{\citenamefont {Abreu}\ \emph {et~al.}(2017)\citenamefont {Abreu}
  \emph {et~al.}}]{Abreu:2017bpe}%
  \BibitemOpen
  \bibfield  {author} {\bibinfo {author} {\bibfnamefont {Y.}~\bibnamefont
  {Abreu}} \emph {et~al.} (\bibinfo {collaboration} {SoLid}),\ }\href {\doibase
  10.1088/1748-0221/12/04/P04024} {\bibfield  {journal} {\bibinfo  {journal}
  {JINST}\ }\textbf {\bibinfo {volume} {12}},\ \bibinfo {pages} {P04024}
  (\bibinfo {year} {2017})},\ \Eprint {http://arxiv.org/abs/1703.01683}
  {arXiv:1703.01683 [physics.ins-det]} \BibitemShut {NoStop}%
\bibitem [{\citenamefont {Ashenfelter}\ \emph
  {et~al.}(2019{\natexlab{a}})\citenamefont {Ashenfelter} \emph
  {et~al.}}]{Ashenfelter:2018zdm}%
  \BibitemOpen
  \bibfield  {author} {\bibinfo {author} {\bibfnamefont {J.}~\bibnamefont
  {Ashenfelter}} \emph {et~al.} (\bibinfo {collaboration} {PROSPECT}),\ }\href
  {\doibase 10.1016/j.nima.2018.12.079} {\bibfield  {journal} {\bibinfo
  {journal} {Nucl. Instrum. Meth.}\ }\textbf {\bibinfo {volume} {A922}},\
  \bibinfo {pages} {287} (\bibinfo {year} {2019}{\natexlab{a}})},\ \Eprint
  {http://arxiv.org/abs/1808.00097} {arXiv:1808.00097 [physics.ins-det]}
  \BibitemShut {NoStop}%
\bibitem [{\citenamefont {Ashenfelter}\ \emph
  {et~al.}(2018{\natexlab{a}})\citenamefont {Ashenfelter} \emph
  {et~al.}}]{Ashenfelter:2018iov}%
  \BibitemOpen
  \bibfield  {author} {\bibinfo {author} {\bibfnamefont {J.}~\bibnamefont
  {Ashenfelter}} \emph {et~al.} (\bibinfo {collaboration} {PROSPECT}),\ }\href
  {\doibase 10.1103/PhysRevLett.121.251802} {\bibfield  {journal} {\bibinfo
  {journal} {Phys. Rev. Lett.}\ }\textbf {\bibinfo {volume} {121}},\ \bibinfo
  {pages} {251802} (\bibinfo {year} {2018}{\natexlab{a}})},\ \Eprint
  {http://arxiv.org/abs/1806.02784} {arXiv:1806.02784 [hep-ex]} \BibitemShut
  {NoStop}%
\bibitem [{\citenamefont {Kwon}\ \emph {et~al.}(1981)\citenamefont {Kwon},
  \citenamefont {Boehm}, \citenamefont {Hahn}, \citenamefont {Henrikson},
  \citenamefont {Vuilleumier}, \citenamefont {Cavaignac}, \citenamefont
  {Koang}, \citenamefont {Vignon}, \citenamefont {Von~Feilitzsch},\ and\
  \citenamefont {Mossbauer}}]{Kwon:1981ua}%
  \BibitemOpen
  \bibfield  {author} {\bibinfo {author} {\bibfnamefont {H.}~\bibnamefont
  {Kwon}}, \bibinfo {author} {\bibfnamefont {F.}~\bibnamefont {Boehm}},
  \bibinfo {author} {\bibfnamefont {A.~A.}\ \bibnamefont {Hahn}}, \bibinfo
  {author} {\bibfnamefont {H.~E.}\ \bibnamefont {Henrikson}}, \bibinfo {author}
  {\bibfnamefont {J.~L.}\ \bibnamefont {Vuilleumier}}, \bibinfo {author}
  {\bibfnamefont {J.~F.}\ \bibnamefont {Cavaignac}}, \bibinfo {author}
  {\bibfnamefont {D.~H.}\ \bibnamefont {Koang}}, \bibinfo {author}
  {\bibfnamefont {B.}~\bibnamefont {Vignon}}, \bibinfo {author} {\bibfnamefont
  {F.}~\bibnamefont {Von~Feilitzsch}}, \ and\ \bibinfo {author} {\bibfnamefont
  {R.~L.}\ \bibnamefont {Mossbauer}},\ }\href {\doibase
  10.1103/PhysRevD.24.1097} {\bibfield  {journal} {\bibinfo  {journal} {Phys.
  Rev.}\ }\textbf {\bibinfo {volume} {D24}},\ \bibinfo {pages} {1097} (\bibinfo
  {year} {1981})}\BibitemShut {NoStop}%
\bibitem [{\citenamefont {Ashenfelter}\ \emph
  {et~al.}(2018{\natexlab{b}})\citenamefont {Ashenfelter} \emph
  {et~al.}}]{Ashenfelter:2018cli}%
  \BibitemOpen
  \bibfield  {author} {\bibinfo {author} {\bibfnamefont {J.}~\bibnamefont
  {Ashenfelter}} \emph {et~al.} (\bibinfo {collaboration} {PROSPECT}),\ }\href
  {\doibase 10.1088/1748-0221/13/06/P06023} {\bibfield  {journal} {\bibinfo
  {journal} {JINST}\ }\textbf {\bibinfo {volume} {13}},\ \bibinfo {pages}
  {P06023} (\bibinfo {year} {2018}{\natexlab{b}})},\ \Eprint
  {http://arxiv.org/abs/1805.09245} {arXiv:1805.09245 [physics.ins-det]}
  \BibitemShut {NoStop}%
\bibitem [{\citenamefont {Ashenfelter}\ \emph
  {et~al.}(2019{\natexlab{b}})\citenamefont {Ashenfelter} \emph
  {et~al.}}]{Ashenfelter:2019iqj}%
  \BibitemOpen
  \bibfield  {author} {\bibinfo {author} {\bibfnamefont {J.}~\bibnamefont
  {Ashenfelter}} \emph {et~al.} (\bibinfo {collaboration} {PROSPECT}),\
  }\href@noop {} {\  (\bibinfo {year} {2019}{\natexlab{b}})},\ \Eprint
  {http://arxiv.org/abs/1901.05569} {arXiv:1901.05569 [physics.ins-det]}
  \BibitemShut {NoStop}%
\bibitem [{\citenamefont {Ashenfelter}\ \emph
  {et~al.}(2019{\natexlab{c}})\citenamefont {Ashenfelter} \emph
  {et~al.}}]{Ashenfelter:2019lbf}%
  \BibitemOpen
  \bibfield  {author} {\bibinfo {author} {\bibfnamefont {J.}~\bibnamefont
  {Ashenfelter}} \emph {et~al.} (\bibinfo {collaboration} {PROSPECT}),\
  }\href@noop {} {\  (\bibinfo {year} {2019}{\natexlab{c}})},\ \Eprint
  {http://arxiv.org/abs/1902.06430} {arXiv:1902.06430 [physics.ins-det]}
  \BibitemShut {NoStop}%
\bibitem [{\citenamefont {Allison}\ \emph {et~al.}(2016)\citenamefont {Allison}
  \emph {et~al.}}]{Allison:2016lfl}%
  \BibitemOpen
  \bibfield  {author} {\bibinfo {author} {\bibfnamefont {J.}~\bibnamefont
  {Allison}} \emph {et~al.},\ }\href {\doibase 10.1016/j.nima.2016.06.125}
  {\bibfield  {journal} {\bibinfo  {journal} {Nucl. Instrum. Meth.}\ }\textbf
  {\bibinfo {volume} {A835}},\ \bibinfo {pages} {186} (\bibinfo {year}
  {2016})}\BibitemShut {NoStop}%
\bibitem [{\citenamefont {BIRKS}(1964)}]{Birks:1964zz}%
  \BibitemOpen
  \bibfield  {author} {\bibinfo {author} {\bibfnamefont {J.}~\bibnamefont
  {BIRKS}},\ }in\ \href {\doibase
  https://doi.org/10.1016/B978-0-08-010472-0.50013-6} {\emph {\bibinfo
  {booktitle} {The Theory and Practice of Scintillation Counting}}},\ \bibinfo
  {series and number} {International Series of Monographs in Electronics and
  Instrumentation},\ \bibinfo {editor} {edited by\ \bibinfo {editor}
  {\bibfnamefont {J.}~\bibnamefont {BIRKS}}}\ (\bibinfo  {publisher}
  {Pergamon},\ \bibinfo {year} {1964})\ pp.\ \bibinfo {pages} {269 --
  320}\BibitemShut {NoStop}%
\bibitem [{\citenamefont {Patrignani}\ \emph {et~al.}(2016)\citenamefont
  {Patrignani}, \citenamefont {Group} \emph {et~al.}}]{patrignani2016passage}%
  \BibitemOpen
  \bibfield  {author} {\bibinfo {author} {\bibfnamefont {C.}~\bibnamefont
  {Patrignani}}, \bibinfo {author} {\bibfnamefont {P.~D.}\ \bibnamefont
  {Group}},  \emph {et~al.},\ }\href@noop {} {\bibfield  {journal} {\bibinfo
  {journal} {Chin. Phys. C}\ }\textbf {\bibinfo {volume} {40}},\ \bibinfo
  {pages} {100001} (\bibinfo {year} {2016})}\BibitemShut {NoStop}%
\bibitem [{sup()}]{supplemental}%
  \BibitemOpen
  \href@noop {} {}\bibinfo {note} {Full detector response matrix, covariance
  matrix, non-$^{235}$U antineutrino corrections, and data points can be found
  in Supplemental Material.}\BibitemShut {Stop}%
\bibitem [{\citenamefont {Lockwood}\ and\ \citenamefont
  {Yingst}(1956)}]{Lockwood:1956}%
  \BibitemOpen
  \bibfield  {author} {\bibinfo {author} {\bibfnamefont {J.~A.}\ \bibnamefont
  {Lockwood}}\ and\ \bibinfo {author} {\bibfnamefont {H.~E.}\ \bibnamefont
  {Yingst}},\ }\href {\doibase 10.1103/PhysRev.104.1718} {\bibfield  {journal}
  {\bibinfo  {journal} {Phys. Rev.}\ }\textbf {\bibinfo {volume} {104}},\
  \bibinfo {pages} {1718} (\bibinfo {year} {1956})}\BibitemShut {NoStop}%
\bibitem [{orn()}]{ornl_weather}%
  \BibitemOpen
  \href@noop {} {\enquote {\bibinfo {title} {{Oak Ridge Reservation
  Meteorology}},}\ }\bibinfo {howpublished} {\url{https://metweb.ornl.gov}},\
  \bibinfo {note} {{Accessed: 2018-11-02}}\BibitemShut {NoStop}%
\bibitem [{\citenamefont {Ilas}\ \emph {et~al.}(2015)\citenamefont {Ilas},
  \citenamefont {Chandler}, \citenamefont {Ade}, \citenamefont {Sunny},
  \citenamefont {Betzler},\ and\ \citenamefont {Pinkston}}]{osti_1185903}%
  \BibitemOpen
  \bibfield  {author} {\bibinfo {author} {\bibfnamefont {G.}~\bibnamefont
  {Ilas}}, \bibinfo {author} {\bibfnamefont {D.}~\bibnamefont {Chandler}},
  \bibinfo {author} {\bibfnamefont {B.~J.}\ \bibnamefont {Ade}}, \bibinfo
  {author} {\bibfnamefont {E.~E.}\ \bibnamefont {Sunny}}, \bibinfo {author}
  {\bibfnamefont {B.~R.}\ \bibnamefont {Betzler}}, \ and\ \bibinfo {author}
  {\bibfnamefont {D.}~\bibnamefont {Pinkston}},\ }\href {\doibase
  10.2172/1185903} {\emph {\bibinfo {title} {Modeling and Simulations for the
  High Flux Isotope Reactor Cycle 400}}},\ \bibinfo {type} {Tech. Rep.}\
  \bibinfo {number} {ORNL/TM-2015/36}\ (\bibinfo {year} {2015})\BibitemShut
  {NoStop}%
\bibitem [{\citenamefont {{Oak Ridge National Laboratory}}(2011)}]{SCALE61}%
  \BibitemOpen
  \bibfield  {author} {\bibinfo {author} {\bibnamefont {{Oak Ridge National
  Laboratory}}},\ }\href@noop {} {\emph {\bibinfo {title} {{SCALE: A
  Comprehensive Modeling and Simulation Suite for Nuclear Safety Analysis and
  Design}}}},\ \bibinfo {type} {Tech. Rep.}\ (\bibinfo {year} {2011})\ \bibinfo
  {note} {{ORNL/TM-2005/39}}\BibitemShut {NoStop}%
\bibitem [{\citenamefont {{X-5 Monte Carlo Team}}(2005)}]{MCNP5}%
  \BibitemOpen
  \bibfield  {author} {\bibinfo {author} {\bibnamefont {{X-5 Monte Carlo
  Team}}},\ }\href@noop {} {\emph {\bibinfo {title} {{MCNP - A General Monte
  Carlo N-Particle Transport Code, Version 5}}}},\ \bibinfo {type} {Tech.
  Rep.}\ (\bibinfo  {institution} {Los Alamos National Laboratory},\ \bibinfo
  {year} {2005})\ \bibinfo {note} {{LA-UR-03-1987}}\BibitemShut {NoStop}%
\bibitem [{\citenamefont {Schmidt}\ \emph {et~al.}(1982)\citenamefont
  {Schmidt}, \citenamefont {Hungerford}, \citenamefont {Daniel}, \citenamefont
  {von Egidy}, \citenamefont {Kerr}, \citenamefont {Brissot}, \citenamefont
  {Barreau}, \citenamefont {Borner}, \citenamefont {Hofmeyr},\ and\
  \citenamefont {Lieb}}]{Schmidt:1982zz}%
  \BibitemOpen
  \bibfield  {author} {\bibinfo {author} {\bibfnamefont {H.~H.}\ \bibnamefont
  {Schmidt}}, \bibinfo {author} {\bibfnamefont {P.}~\bibnamefont {Hungerford}},
  \bibinfo {author} {\bibfnamefont {H.}~\bibnamefont {Daniel}}, \bibinfo
  {author} {\bibfnamefont {T.}~\bibnamefont {von Egidy}}, \bibinfo {author}
  {\bibfnamefont {S.~A.}\ \bibnamefont {Kerr}}, \bibinfo {author}
  {\bibfnamefont {R.}~\bibnamefont {Brissot}}, \bibinfo {author} {\bibfnamefont
  {G.}~\bibnamefont {Barreau}}, \bibinfo {author} {\bibfnamefont {H.~G.}\
  \bibnamefont {Borner}}, \bibinfo {author} {\bibfnamefont {C.}~\bibnamefont
  {Hofmeyr}}, \ and\ \bibinfo {author} {\bibfnamefont {K.~P.}\ \bibnamefont
  {Lieb}},\ }\href {\doibase 10.1103/PhysRevC.25.2888} {\bibfield  {journal}
  {\bibinfo  {journal} {Phys. Rev.}\ }\textbf {\bibinfo {volume} {C25}},\
  \bibinfo {pages} {2888} (\bibinfo {year} {1982})}\BibitemShut {NoStop}%
\bibitem [{\citenamefont {Dwyer}(2015)}]{oklo}%
  \BibitemOpen
  \bibfield  {author} {\bibinfo {author} {\bibfnamefont {D.}~\bibnamefont
  {Dwyer}},\ }\href@noop {} {\enquote {\bibinfo {title} {Oklo},}\ }\bibinfo
  {howpublished} {\url{https://github.com/dadwyer/oklo}} (\bibinfo {year}
  {2015})\BibitemShut {NoStop}%
\bibitem [{\citenamefont {An}\ \emph {et~al.}(2017{\natexlab{b}})\citenamefont
  {An} \emph {et~al.}}]{An:2016srz}%
  \BibitemOpen
  \bibfield  {author} {\bibinfo {author} {\bibfnamefont {F.~P.}\ \bibnamefont
  {An}} \emph {et~al.} (\bibinfo {collaboration} {Daya Bay}),\ }\href {\doibase
  10.1088/1674-1137/41/1/013002} {\bibfield  {journal} {\bibinfo  {journal}
  {Chin. Phys.}\ }\textbf {\bibinfo {volume} {C41}},\ \bibinfo {pages} {013002}
  (\bibinfo {year} {2017}{\natexlab{b}})},\ \Eprint
  {http://arxiv.org/abs/1607.05378} {arXiv:1607.05378 [hep-ex]} \BibitemShut
  {NoStop}%
\bibitem [{\citenamefont {Buck}\ \emph {et~al.}(2017)\citenamefont {Buck},
  \citenamefont {Collin}, \citenamefont {Haser},\ and\ \citenamefont
  {Lindner}}]{Buck:2015clx}%
  \BibitemOpen
  \bibfield  {author} {\bibinfo {author} {\bibfnamefont {C.}~\bibnamefont
  {Buck}}, \bibinfo {author} {\bibfnamefont {A.~P.}\ \bibnamefont {Collin}},
  \bibinfo {author} {\bibfnamefont {J.}~\bibnamefont {Haser}}, \ and\ \bibinfo
  {author} {\bibfnamefont {M.}~\bibnamefont {Lindner}},\ }\href {\doibase
  10.1016/j.physletb.2016.11.062} {\bibfield  {journal} {\bibinfo  {journal}
  {Phys. Lett.}\ }\textbf {\bibinfo {volume} {B765}},\ \bibinfo {pages} {159}
  (\bibinfo {year} {2017})},\ \Eprint {http://arxiv.org/abs/1512.06656}
  {arXiv:1512.06656 [hep-ex]} \BibitemShut {NoStop}%
\bibitem [{\citenamefont {Huber}(2017)}]{Huber:2016xis}%
  \BibitemOpen
  \bibfield  {author} {\bibinfo {author} {\bibfnamefont {P.}~\bibnamefont
  {Huber}},\ }\href {\doibase 10.1103/PhysRevLett.118.042502} {\bibfield
  {journal} {\bibinfo  {journal} {Phys. Rev. Lett.}\ }\textbf {\bibinfo
  {volume} {118}},\ \bibinfo {pages} {042502} (\bibinfo {year} {2017})},\
  \Eprint {http://arxiv.org/abs/1609.03910} {arXiv:1609.03910 [hep-ph]}
  \BibitemShut {NoStop}%
\end{thebibliography}%

\end{document}